\documentclass[scrartcl,aps,pra,reprint,superscriptaddress,amsmath,showpacs,amsfonts,amssymb,onecolumn,bibliography,11pt]{revtex4-1}

\usepackage{graphicx}
\usepackage{bm}
\usepackage{color}
\usepackage{MnSymbol}
\usepackage{enumerate}
\usepackage{bbold}
\usepackage{subfigure}
\usepackage{units}
\usepackage{gensymb}

\def\be{\begin{equation}}
\def\ee{\end{equation}}

\begin{document}

\title{Deep Radiative Cooling Passive Dew Collection}

\author{Yuval Zamir}
\author{Nadav Drechsler}
\author{John C. Howell}
\affiliation{Racah Institute of Physics, The Hebrew University of Jerusalem, Jerusalem, Israel, 91904}

\begin{abstract}
We show how to overcome almost every major limitation in passive dew collection including: parasitic heating, parasitic evaporation, uncontrolled wind conditions, and performing dew collection 24 hours per day at nearly arbitrary relative humidity. We incorporate an ideal selective emitter in a simple thermal-equilibrium dew collection model that relies on an important physical insight -- that the roles of the radiative emitter and dew collector should be decoupled to achieve optimal dew harvesting.  Previous models of passive dew collection strongly coupled those roles leading to nonideal theoretical limits.  Our model necessitates deep radiative cooling (i.e., radiative cooling very far below ambient temperature) and that all forms of conductive and convective heating of the emitter have been eliminated.  We further propose and tested a passive, self-regulating, gravity-fed airflow design that eliminates uncontrollable wind conditions or externally-driven airflow systems. Under somewhat realistic atmospheric scenarios, the theory predicts in excess of 80 grams/m$^2$/hour and 1.5 L/m$^2$/day which greatly exceeds the current theoretical limits of 0.7 L/m$^2$/day. We discuss how this can be further improved.  We tested the validity of the equilibrium model and the gravity-fed airflow using a laboratory experiment that replicated the radiative emitter/dew collector system with the hot and cold side of a well-calibrated thermo-electric cooler. These ideas not only point to improved dew collection yields, but that dew can be collected 24 hours per day even at low relative humidity.  
\end{abstract}

\date{\today}
\maketitle

\section{Introduction}
Water stress/scarcity is one of the most urgent problems facing mankind \cite{RisksReport2019}. Mekonnen and Hoekstra showed that over half of the world's population suffer from severe water scarcity at least 1 month of the year \cite{Mekonnene1500323}. The significant concerns of water scarcity and increased demand are driving an intense push in research and development of water production. Of the many research directions, air-water generation plays an important role because water vapor is inherently renewable, local, vast (estimated at about 13,000 km$^3$ \cite{hamed2010technical} in the atmosphere) and potable, albeit likely after some filtration \cite{muselli2006dew}. Among the most historically successful methods employed for air-water generation is dew collection \cite{AGAM2006572,Khalil2016}. Passive methods for dew collection, exploiting radiative cooling \cite{Hossain2016,ZEYGHAMI2018115} are of great value , because of the ease of use and because they do not require any external power source. However, maximum yields on passive devices have been low, typically in the range of 200 mL/m$^2$ to 600 mL/m$^2$ per day \cite{JACOBS2008377,muselli2009dew,maestre2011comparative,lekouch2012rooftop,MAESTREVALERO2012103,BEYSENS2016146}. 

While there are a host of parameters that can modify yields including: materials \cite{maestre2011comparative,ZEYGHAMI2018115}, shape \cite{JACOBS2008377,clus2009comparison}, size, orientation, wind conditions \cite{beysens2003using,gandhidasan2005modeling,clus2008study,muselli2009dew}, cloud cover \cite{Khalil2016}, hydrophobicity \cite{seo2016effects} etc. by and large dew collectors have collected dew on top of the radiative cooling mechanism.  

In a seminal book by Monteith and Unsworth \cite{monteith2013principles} they used energy balance equations to set upper limits on dew-on-emitter system collection yields at about 800 mL/m$^2$/day, which has been considered the upper limit for decades. Recently, Beysens revisited these theoretical upper limits by considering only a few important meteorological parameters. He performed experiments using 10 dew collectors distributed throughout the world along with a carefully controlled laboratory setup to arrive at an estimated maximum dew collection of 700 mL/m$^2$/day \cite{BEYSENS2016146}. Beysen's results assumed a planar, dew-on-emitter, surface exposed to laminar air flow with between 11 and 12 hours per day of possible dew collection.  

Disadvantages of dew-on-emitter systems include: shallow cooling (cooling only slightly below the ambient), only work in nighttime conditions, are limited to moderately high to high relative humidity, suffer from parasitic processes \cite{BEYSENS2016146}, are often subject to uncontrollable wind conditions, and are subject to dust and environmental pollutants \cite {Muselli2002}. The nighttime only and shallow cooling characteristics are from the absorption/emission properties of water.  As soon as dew forms on the emitter, the emitter can take on the broadband emission properties of water and destroy the selective emission. Broadband emitters cannot achieve deep cooling \cite{,Hossain2016,ZEYGHAMI2018115}.  The broadband water absorption/emission extends to the near infra-red (IR) meaning that dew-on-emitter systems cannot be used during daylight, since there is significant heating from the Sun in the near-IR.  

\begin{figure}[tb!]
\centering \includegraphics[width=.8\linewidth]{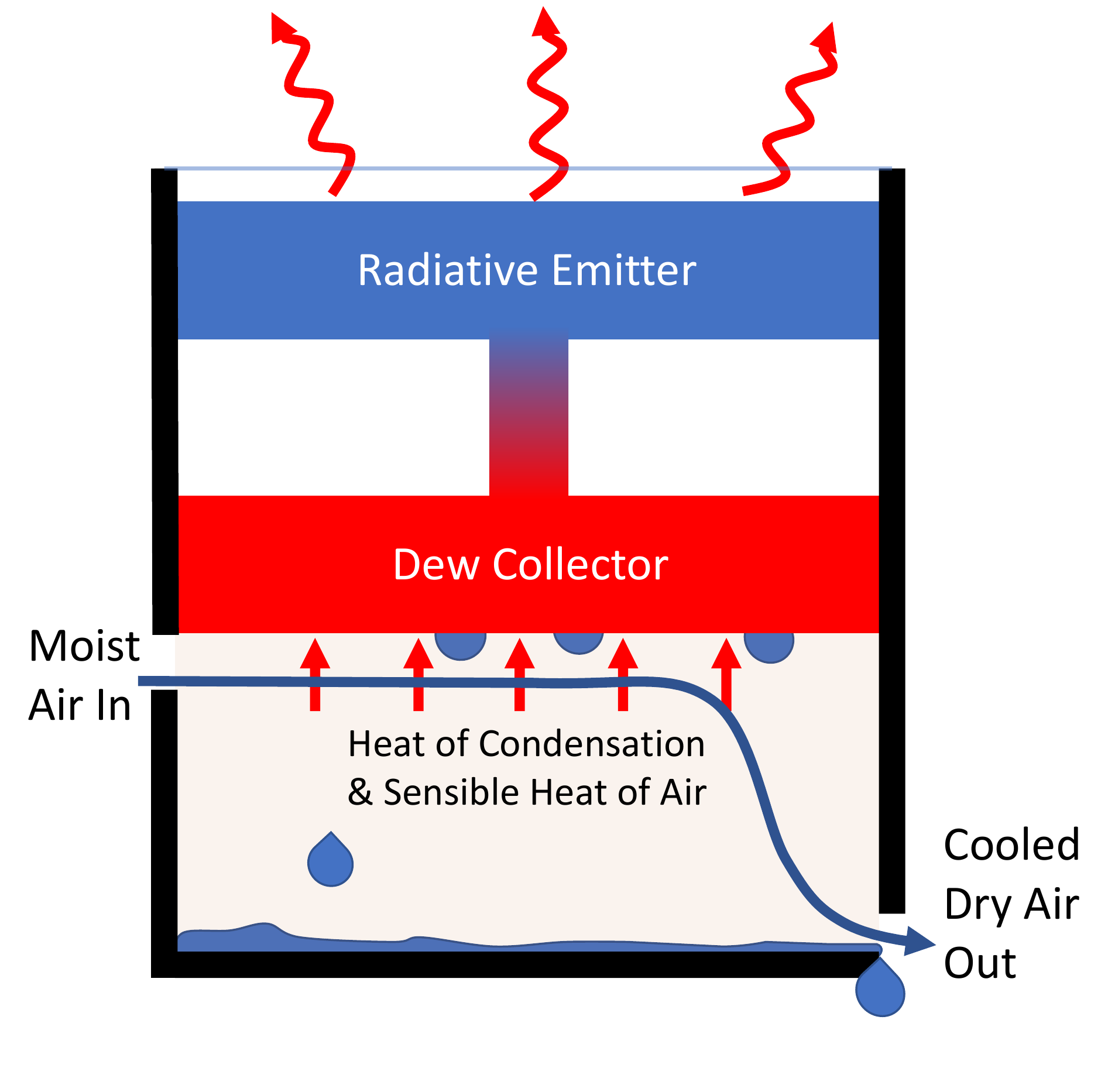}
\caption{A generalized decoupled dew collection system. The radiative emitter and the dew collector are in distinct locations with heat transferred between the dew collector and the radiative emitter. The emitter properties are solely affected by the heat coming from the collector and not from the dew process. A passive, gravity-fed, air flow system is shown. Warm air enters the system and as the air cools and dries, it falls.}
\hfill{}\null \label{DewCollectionSchematic} 
\end{figure}

Inspired by recent developments in deep radiative cooling \cite{chen2016radiative,zhou2018accelerating}, we reconsider theoretical limits of systems when the roles of the radiative emitter and the dew collector are decoupled. Deep radiative cooling is made possible by creating devices that strongly emit in the atmospheric transmission, but weakly otherwise.   The decoupling allows the emitter to achieve and, importantly, to maintain deep radiative cooling far below the dew point even during the dew collection process.  A decoupled design is pictorially shown in Fig. \ref{DewCollectionSchematic}. A radiative emitter cools as it undergoes heat exchange with space through the spectral atmospheric transmission window. The radiative emitter and dew collection system undergo efficient heat exchange with each other, but perform distinct roles. The dew collection system removes ``sensible" heat from the air as well as  the latent heat of condensation as the vapor condenses on the collector. We can nearly minimize heat losses and uncontrolled effects of wind by using a gravity-fed airflow system in a chamber with saturated air.  While we arrive at approximately the same theoretical thermodynamic upper bounds as Beysens and others under ideal conditions (i.e., clear sky, nighttime, high relative humidity), we show that our system can achieve much higher yields in low relative humidity and during daylight hours.

\section{Theory}
We now consider a thermal equilibrium theory model for a decoupled radiative emitter/dew collector system.  We assume the use of a deep radiative cooler, which requires a highly selective radiative emitter having cooling power per unit area
\begin{equation}
    P_{cool}(T_e,T_{amb})=P_{rad}(T_e)-P_{atm}(T_{amb})
\end{equation}
given in W/m$^2$.  We assume the use of a selective emitter that has unit emissivity between 8 and 13 microns and zero emissivity otherwise. We approximate the radiated power per meter squared to be $P_{rad}=a\sigma T_e^4$, where $\sigma$ is the Stefan-Boltzmann constant, $T_e$ is the radiator temperature and $a$ is an emitter-dependent parameter that incorporates the spectral and geometrical emission properties of the emitter. For a given range of temperatures, a good value for $a$ can be found by integrating the blackbody spectrum over the atmospheric transmission window as a ratio with the full integration over the black body spectrum. For temperature ranges of interest in this work, we will use $a=0.326$.  $P_{atm}(T_{amb})=b\sigma T_{amb}^4$ is the atmospheric radiated power absorbed by the emitter per meter squared.  The value of $b$ is found by integrating over the angular and wavelength properties of the atmosphere. We make the assumption that the atmosphere behaves as a grey body of uniform emissivity between 8 and 13 microns with an emissivity of approximately 24\%. In making this assumption, we assume that the emitter geometry is such that it eliminates most of the large angle (with respect to the zenith) radiation coming from the atmosphere.  This means that the atmospheric emission will be roughly constant and close to zenith emissivity.  Once again, for the temperature ranges of the atmosphere we will explore here, we assume a value of $b=0.080$.

The values of $a$ and $b$ are temperature-dependent, but the variation is small assuming the temperature range is much smaller than the temperature. Physically speaking, this is a good assumption for two reasons.  First, the spectrum of the blackbody does not shift much for small temperature changes, but the magnitude of the spectral irradiance does.  Second, since the peak of the blackbody spectrum of either the emitter or ambient is wide and the peak of the spectrum falls between the endpoints of the transmission window, small movements don't change the integration significantly.  

Additionally, we assume no other forms of heating including from the Sun or from non-radiative sources such as conduction or convection.  For an ambient temperature of 20C, these approximations predict a radiative cooling power of approximately 100 W/m$^2$ when the radiator is at ambient temperature.  

Our goal was to develop a simple, analytical thermal equilibrium model from first principles. It is similar to many other models such as those developed by Beysens \cite{BEYSENS2016146} and to the energy model used by Awanou, Hazoume and  Kounouhewa \cite{Awanou1997, Awanou1999}. Owing to a passive, gravity-fed airflow design with relative slow velocities, we do not need to worry about nonlinear convective heat flow. Suppose that during a unit of time, a volume of air $V$ is brought into a chamber. The chamber is in thermal equilibrium with the emitter. The air is cooled until it is in thermal equilibrium with the chamber at temperature $T$, hence $T_e=T$. If the air temperature goes below the dew point $T_{d}$, condensation forms on the surface of the chamber. The total energy cost is $E_{a}(V,T)+E_{c}(V,T)$ where $E_{a}(V,T)$ is the energy needed to cool the air and $E_{c}(V,T)$ is the energy needed for the heat of condensation. Here, 
\begin{equation}
E_{a}(V,T)=C_{p}\rho V(T_{i}-T),
\end{equation}
where $C_{p}=1005$ J/kgK is the heat capacity of dry air at constant pressure, $\rho=1.23$ kg/m$^3$ is the density of the air, $T_{i}$ is the initial temperature (of the ambient) and $T$ is the final temperature of the air (or the emitter) as described earlier. As a note, there is an absolute humidity-dependent correction to the heat capacity for air, but it is small (of the order of a few percent) for the absolute humidities we will explore here.

To determine the energy consumed during the condensation process, we need to know the mass of the water condensed and multiply it by the heat of condensation per unit mass, which is approximately 2.5 MJ/kg at the freezing point and is approximately 2.26 MJ/kg at the boiling point having approximately a linear decrease with temperature. Throughout the remainder of the paper, we will assume that the heat of condensation has the linear relationship to temperature 
\begin{equation}
L(T)=-2400(T-273)+2.5\cdot10^{6}
\end{equation}
in units of J/kg. It is assumed that the water condenses at the temperature $T$ of the emitter. 

The total mass of water condensed is the difference in the mass of water in the air at the dew point and the mass of water in the saturated air at temperature $T$. Assuming the water vapor behave as an ideal gas, the mass of water in saturated air of volume $V$ at temperature $T$ is given by 
\begin{equation}
m(V,T)=\frac{M}{1000}\cdot \frac{p_{weq}(T)V}{RT}
\end{equation}
in units of kg, where 
\begin{equation}
p_{weq}(T)=0.0317\cdot10^{5}\exp\left[\frac{L}{R}\cdot\left(\frac{1}{298}-\frac{1}{T}\right)\right]
\end{equation} 
is the equilibrium vapor pressure in units of Pa \cite{Cluasius,VapourPressureTable}, the mass is given in kg, M is the molar mass of water (18.015), R is the gas constant. 

Therefore, the mass of water extracted per unit volume of processed air as a function of the final temperature T, is given by:
\begin{equation}
\Delta\hat{m}(T)=\frac{M}{1000}\cdot \frac{p_{weq}(T_d)-p_{weq}(T)}{RT}.
\end{equation}

We make the simplifying mathematical assumption that the vapor-liquid water phase transition occurs at the temperature of the emitter. Then, the energy needed to condense the water to temperature T, (which only happens when $T<T_{d}$)
is 
\begin{equation}
E_{c}(V,T)=L(T)(m(V,T_{d})-m(V,T))=L(T)\cdot \Delta\hat{m}(T)\cdot V.
\end{equation}
The alternative is to assume that the vapor continuously condenses from the dew point down to the temperature of the emitter, but needing additional energy to cool the liquid water down to the emitter temperature. Owing to the higher heat of condensation at lower temperatures, these two alternatives only differ by at most, and usually much less than, a percent for most dew collecting conditions.

In thermal equilibrium, with no other parasitic terms, we arrive at the energy balance equation 
\begin{equation}
E_{cool}(T,T_{amb})=E_{a}(V,T)+E_{c}(V,T),
\label{energy}
\end{equation}
where $E_{cool}$ is the time-integrated radiated cooling power of the emitter. This equation sets condensation yield as a function of temperature for a given relative humidity. From above, it can be seen that $E_{c}$ and $E_{a}$ are both linear in the volume $V$ of the air that is in the system. By knowing the required energy to process a unit volume of air, we can find the processed volume for a given radiative energy, $E_{cool}(T)$:
\begin{equation}
V=\frac{E_{cool}(T,T_{amb})}{\hat{E}_{a}(T)+\hat{E}_{c}(T)},
\end{equation}
where $\hat{E}_{a}(T),\hat{E}_{c}(T)$ are the corresponding energy expressions per unit volume.

Finally, we can obtain the amount of water collected for a given dew point $T_d$ as a function of T:
\begin{equation}
\Delta m(T,T_{amb})=\Delta\hat{m}(T)\cdot V(T,,T_{amb})=\Delta\hat{m}(T)\cdot\frac{E_{cool}(T)}{\hat{E}_{a}(T)+\hat{E}_{c}(T)}
\label{mass_limit}
\end{equation}

There are important tradeoffs to consider. At high flow rates, more water enters the system, which requires more air to be cooled resulting in higher equilibrium temperatures. If too much air flows, it may not be possible to cool the air to the dew point. At low flow rates, the temperature of the emitter drops, but the exponential drop of the partial pressure of the water vapor as a function of temperature results in diminishing yield. Further, at low temperatures the radiative cooling power drops. These tradeoffs imply that optimal flow rates exist. 

\section{Analysis and Results}
Now let us consider a numerical example. We next assume a dew point temperature of 291K with four unique ambient temperatures of 292K, 297K, 302K and 307K. The relative humidity of these four temperatures is 94\%, 69\%, 51\% and 39\%, respectively. The dew point is the same for all these temperatures. This implies that the atmospheric water column is approximately the same meaning that the approximations for $a$ and $b$ are reasonable over the range of temperatures explored.  

\begin{figure}[ht!]
  \centering  \includegraphics[width=1\linewidth]{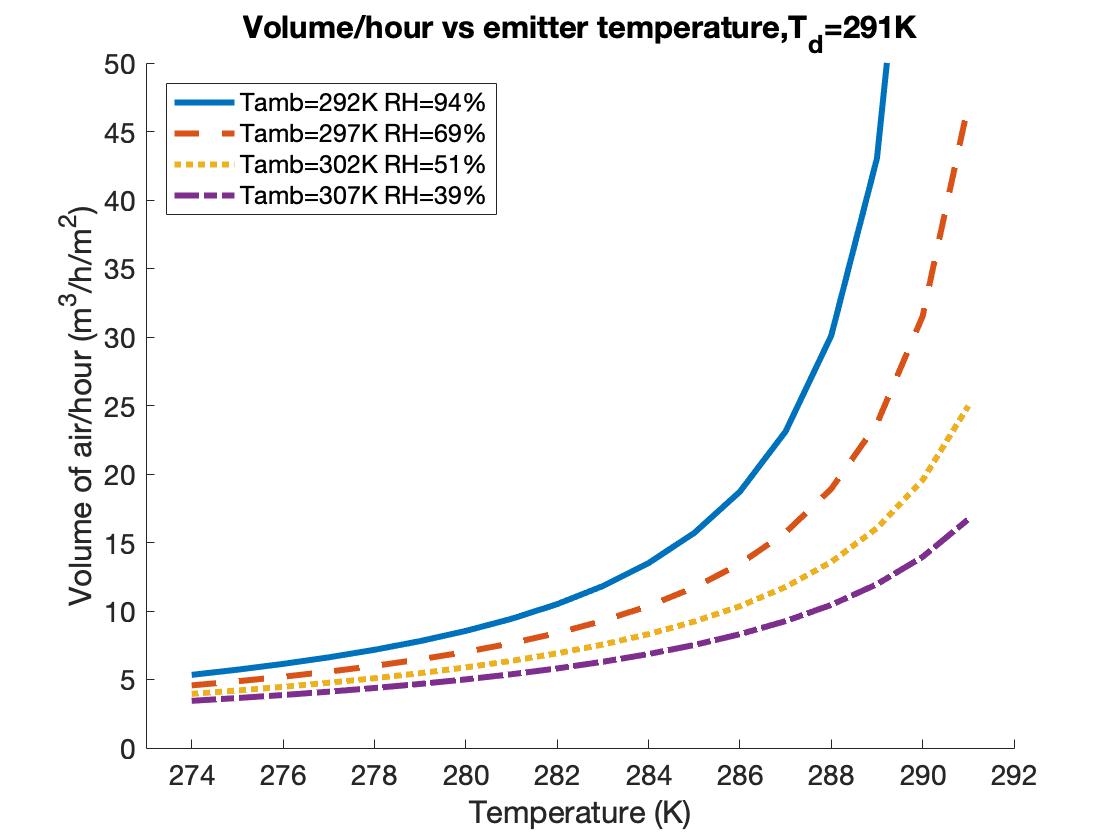}
  \caption{Volume of processed air vs equilibrium temperature. A volume $V$ of air per hour is in thermal equilibrium with a radiative emitter are brought into thermal equilibrium resulting in the equilibrium temperature $T$. The volume of air that is cooled by the emitter vs emitter temperature for the cooled air given an ambient temperature. The dew point temperature is 291K.  The curves show the volumetric rate versus temperature for four different ambient temperatures ranging from 293K to 308K in increments of 5 degrees.}
  \hfill\null
  \label{300Ambient293DewVolume}
\end{figure}

We now ask what is the volume of air that can be cooled for a given equilibrium temperature of the emitter-air system and a given cooling power. The volume versus equilibrium temperature is shown in Fig. \ref{300Ambient293DewVolume}. As one would expect, the more air that is processed by the system, the higher the equilibrium temperature will be for the air-emitter system. 

\begin{figure}[ht!]
 \centering  \includegraphics[width=1\linewidth]{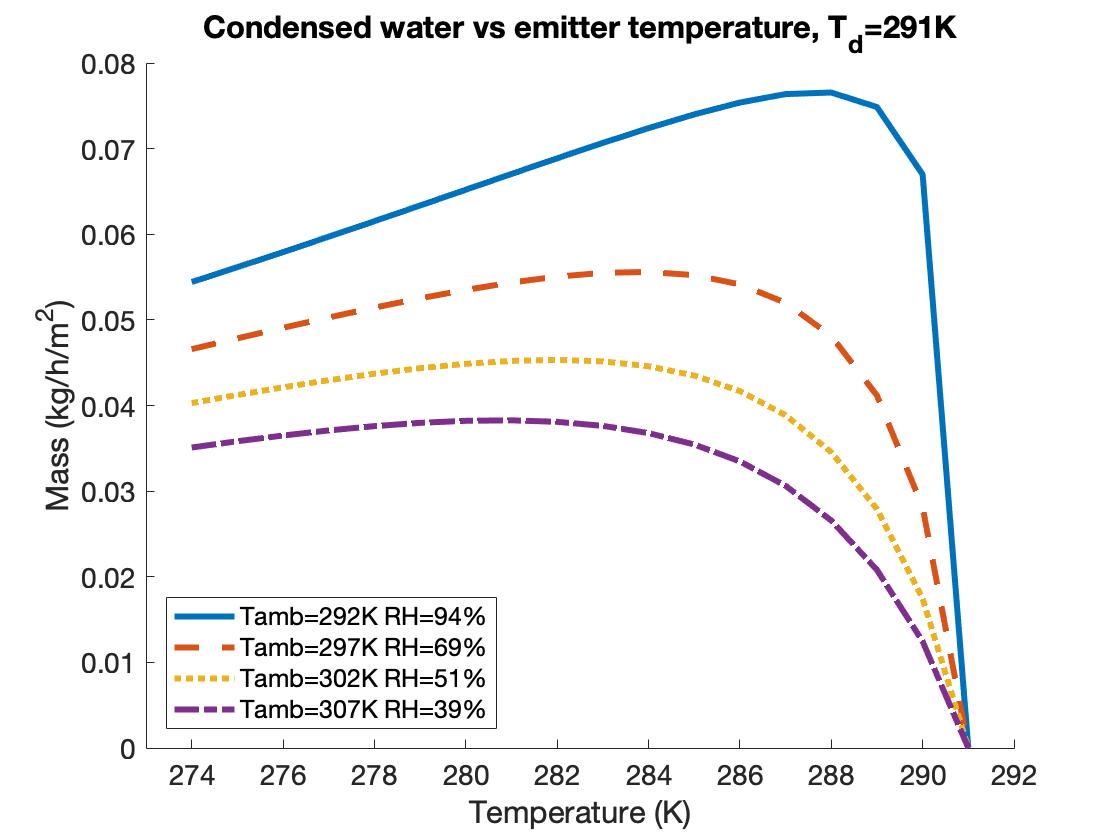}
  \caption{Mass of condensed water vs equilibrium temperature. This shows the cooled air for an ambient temperature ranging from 293K to 308K with a dew point temperature of 291K. }
  \hfill\null
  \label{300Ambient293DewMass}
\end{figure}

The mass of condensed water vs equilibrium temperature is shown in Fig. \ref{300Ambient293DewMass}.  One can see that with 94\% relative humidity, the peak water collected is approximately 77 grams/m$^2$/hour.  Then for a 12 hour period, this would result in over 900 grams of accumulated water.  Further calculations showed that keeping the same dew point, but having a relative humidity of 99\% could yield a maximum dew collection of 87 grams/m$^2$/hour.  If we use the same water collection curves in Fig.  \ref{300Ambient293DewMass} and assume that the dew point remains the same during the day and night but the ambient temperature changes such that the relative humidity drops to 51\% for 12 hours, then a 24 hour dew collection can yield approximately 1.5 L/m$^2$/day.      

One of the most important features of these figures is that the mass of condensed water has only a relatively moderate  variation over a fairly wide range of temperatures and processed volume rates for low to moderately high relative humidity for temperatures well below the dew point. This is intriguing, because it means we can work in a wide range of flow rates without significantly jeopardizing the performance of the system. 

It is also important to note that in the ideal case of no non-radiative losses, knowing the input temperature, the relative humidity of the air, the power of the radiative cooling system, and the emitter temperature, one can uniquely determine the condensed water mass without knowing the airflow. This is true because the information about the airflow is contained in the temperature of the emitter. If more air flows past the emitter, the temperature of the emitter will increase. These facts allow us to use a passive, self-regulating air flow system based on the relative densities of the input and cooled/dried air after interacting with the emitter. 

Another interesting feature in these graphs are that in high relative humidity environments, one prefers to have a relatively high flow rate, whereas in low humidity environments, it is optimal to squeeze as much water as possible from a given volume, since the peak water collection moves to lower temperatures.  For dew-on-emitter systems, the parasitic heating and evaporation don't even allow for low relative humidity dew collection. \cite{BEYSENS2016146, JACOBS2008377}.  Most importantly, the graphs point to dew harvesting yields greatly exceeding current limits.

\section{Experiment: Passive Air Regulation and Dew Collection}
\begin{figure}[h!]
\centering \includegraphics[width=1\linewidth]{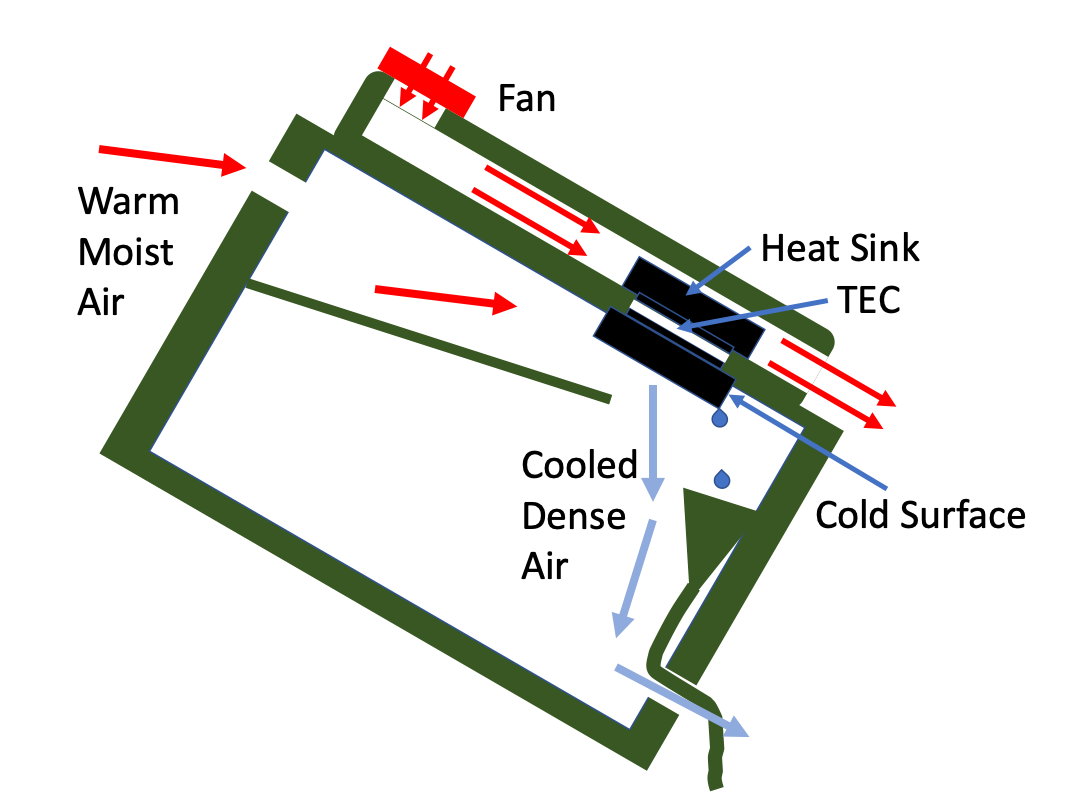}
\caption{Experimental setup. A thermo-electric cooling system simulates a condensation-below-emitter dew collection system. A fan blows away hot air from the hot side of the TEC in analogy to radiative emission of heat. For the dew collection, warm air enters a chamber, interacts with cold aluminum fins, connected to the cold side of the thermo-electic cooler, which condense the water from the air. The cooled and dry air was more dense than the input air and thus fell and left the system through the holes in the bottom. The liquid water dripped into a funnel and was collected in a container outside the system on a sensitive scale with real-time sampling.}
\hfill{}\null \label{DewCollectionSetup} 
\end{figure}

Since we have decoupled the roles of radiative cooling and the dew collection and chen \textit{et al} \cite{chen2016radiative} have already demonstrated the deep radiative emitter, we demonstrate the validity of our model and the passive airflow system.  Instead of using a radiative emitter, we used a proof-of-principle thermoelectric cooling (TEC) system that replicated the properties of a radiative cooler (as is shown in Fig. \ref{DewCollectionSetup}), much as Beysens did to test upper theoretical bounds \cite{BEYSENS2016146}. The TEC is a very good system in which to test the theory without requiring state-of-the-art radiative cooling. The experiment tested the passive, self-regulating air flow system. Aluminum fins are placed both on the hot and cold side of a TEC. A fan blows ambient air through the hot fins to take away the heat akin to radiative emission of heat. It is important to note that this is a dew collection simulator and that while the fan blows away heat, in the real system that heat would have been radiated away by the emitter.  However, no fan is used to drive air into the cooling system to assure it remains fully passive.  The TEC is run at constant power. We ran the TEC system well below its maximum cooling power. 

While the input power of the TEC was constant during the experiment, the cooling power of the TEC is still dependent on the relative temperature of the hot and cold side of the TEC $\Delta T$. Before running the dew collection experiment, we carefully calibrated the cooling power of the TEC as a function of the relative temperature by monitoring $\Delta T$ when the fins are immersed in a precisely known mass of water both with the TEC on and off. For a given constant input voltage (3.8V), we found a linear cooling power function of the temperature difference between the hot and cold sides of the TEC of $P_{TEC}(\Delta T) = -0.527\Delta T + 11.94$ measured in Watts in good agreement with the manufacturer's specifications.

For the passive air flow, we made four 3cm diameter apertures near the top of the system on one side and four holes of the same size near the bottom on the other side. We tilted the system to increase the height differential between the apertures as well as to a create preferential drip direction for the condensed water. When the system is tilted at about 35 degrees from the horizontal, the orientation of the apertures relative to each other is approximately 60 degrees from the horizontal. We added a shelf to direct the airflow to the cold surface. The condensed water drips into a funnel and is collected in a chamber outside the system on a scale with 100 $\mu g$ sensitivity with real-time sampling.  

Warm moist air enters the upper apertures. The air propagates towards the cooled aluminum fins. The air cools and water vapor condenses on the fins. The cooled drier air is more dense than the input air and thus falls in the container towards the lower apertures and leaves the system. To within certain limits, the airflow system is self-regulating because if the fins start to cool, the air becomes colder and more air flows, which heats up the cooled fins and decreases the airflow. To regulate the airflow, one can increase or decrease the size or number of apertures. We believe the gravity-fed design also allows the system to well approximate the thermal equilibrium model.  The air that is the coldest and driest (in thermal equilibrium), leaves the system more rapidly than the warm moist air that will continue to interact with the fins.

\section{Results}

We ran the experiment under several different environmental conditions. By measuring the ambient temperature, relative humidity, and the temperature differential between the hot and cold sides of the TEC through time, we could calculate the theoretical dew collection at each point in time based on the energy balance equation, (\ref{mass_limit}). We compared these calculations with the amount of water collected on the scale. 

\begin{figure}[ht!]
\includegraphics[width=1\linewidth]{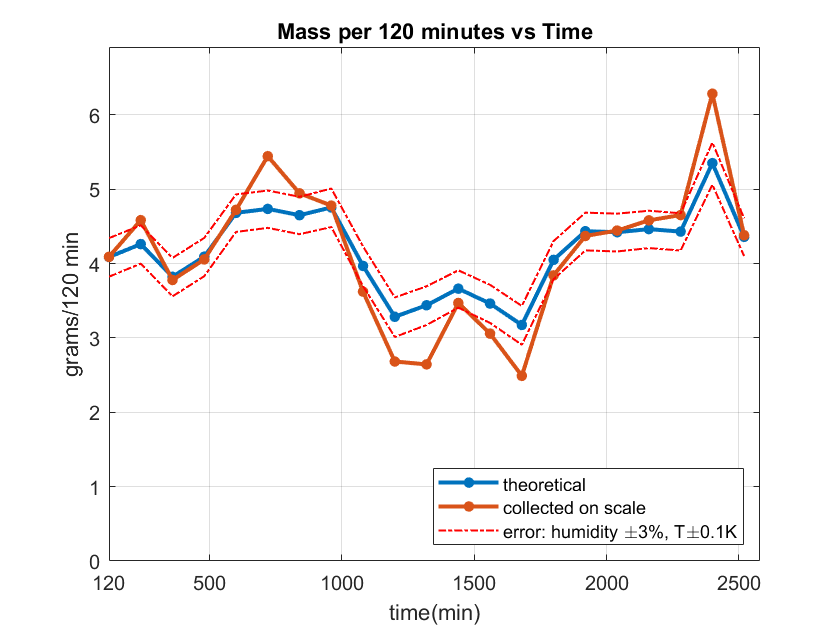}
\caption{Water collected vs time with an average of 2.8W of cooling power. The theoretical limit through time was calculated by measuring the ambient temperature and humidity and the power of the system. This limit is compared with the actual water collected on the scale through time.}
\hfill{}\null \label{WaterVsTime} 
\end{figure}

For the first experiment, shown in Fig. \ref{WaterVsTime}, the system ran for about 2580 minutes, with an average cooling power of 2.8 watts. The average temperature difference between the hot and cold side was 17.2C with 0.66C standard deviation. The average humidity was 73\% with 4.5\% standard deviation. The average ambient temperature was 297.8K with 0.6K standard deviation. The fluctuations over time in the predicted yield shown are in Fig. \ref{WaterVsTime} are more a function of the strong dependence of the cooling power of the TEC as a function of $\Delta T$, which is itself dependent on ambient temperature and humidity. Owing to the fact that at time 0, no water had formed, we chose to start the acquired mass after 100 minutes running to give the system time to come to equilibrium. 

We compiled the results from several other data sets to determine the capabilities of the system under several scenarios.  When addressing the entire data, one may conclude a typical ratio between the distance from  dew point ($T_{ambient} - T_{dew}$) and  the water yield ($\frac{gr}{hour}$): $Yield = -0.81\Delta T + 8.31$. It is important to notice that this formula is typical of this specific system, and is not a general characteristic of the model as a whole. The linearity of this ratio is within agreement with the slope depicted in Fig \ref{300Ambient293DewMass}, under the approximation where $T_{ambient} - T_{cold} = {constant}$.  

\begin{figure}[ht]
\includegraphics[width=1\linewidth]{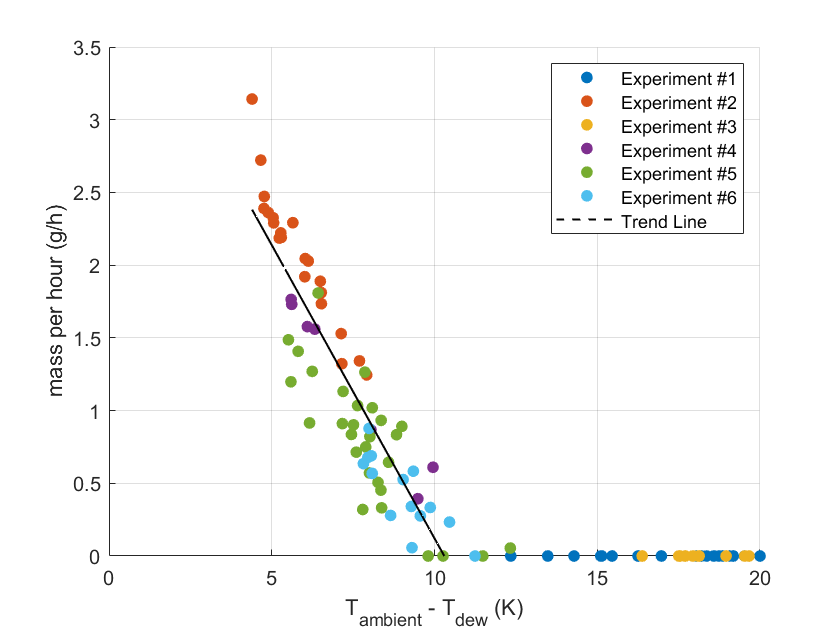}
\caption{Water collected per hour vs. the average difference between ambient temperature and dew point during this hour, from different experiments with different environmental conditions}
\hfill{}\null \label{WaterVsdT} 
\end{figure}

There are a couple factors at play in Fig. \ref{WaterVsdT}.  First, as the temperature difference between the ambient and dew point increases, the cooling power of the TEC decreases as discussed earlier.  Second, we have made the assumption in the theory that the chamber contains fully saturated air at the emitter temperature.  However, this is not the case in the experiment, because we measured non-zero cooling power at points in which there was no dew formed.  While it improves on existing systems, the proposed system is still not yet ideal.    

\section{Discussion}
As discussed earlier, we believe that the system, under the right atmospheric conditions, can achieve upwards of 1.5 L/$m^2$/day.  We note that 73\% relative humidity is at or below the threshold of dew collection for previous radiatively cooled systems \cite{muselli2006comparative,lekouch2012rooftop}. So, not only can the system greatly exceed the maximum yield, it can do it at a relative humidity at which other systems wouldn't collect any dew at all. This property was demonstrated with the production of water even down to almost 50\% relative humidity. These results are possible because of the deep cooling made by using a selective emitter and decoupling the roles of the radiative cooling from dew collection.   

The approximations in this work provide reasonable estimates to radiative cooling power of a generic system at temperatures near those described.  However, to truly get an accurate estimate of the cooling power of a radiative system requires knowing the spectral emission properties of the emitter, the ambient temperature, relative humidity, altitude, the height of the water column in the atmosphere, the local CO$_2$ levels to determine the atmospheric spectral downwelling radiation etc..  Atmospheric modeling software such ModTran are commonly available to calculate such results.  Every locale would then have its specific capabilities.     

The dew collection chamber allows for a low parasitic evaporation environment. Inside the chamber, the air is cooler and in an ideal situation has saturated air.  This means that there will be very little parasitic evaporation. The experimental system just described improved upon this concept, by producing water in conditions where typical dew-on-emitter system would have suffered greatly from re-vaporization of dew. Yet, we believe that the current design - specifically the relatively large condensation chamber and many apertures - still does not completely prevent parasitic evaporation. Thus, we estimate that the low water yield in dry conditions is at least partly caused by the re-vaporization of condensed droplets, even within the condensation chamber. It is reasonable to assume that further research dedicated to optimizing the air flow and geometry will allow to significantly decrease the affect of this issue.

By decoupling the roles of the emitter and collector, the dew collection system can be hidden from the evaporative processes associated with sunlight and the convective heating from exposure to the ambient air.  This allows the system to work 24 hours per day unlike dew-on-emitter systems. In addition, because water has very strong absorption/emission properties in much of the spectrum, the emission properties of the radiator are not affected if they are collected in another location.    

While not done in this work, additional dew can be harvested by recycling the cooled air.  The theory stated that the sensible heat and latent were removed from the air.  Significant amounts of energy were used to remove the sensible heat.  In the system we developed, the cooled air is just dumped back into the environment.  But, it can be further used to cool an element exposed to the atmosphere which could extract even more dew.  

Lastly, and perhaps, with an eye to the future, we have described a decoupled emitter-collector system in which dew is collected on the bottom of the emitter, but does not affect the emitter properties. However, no such local limitation is required. The emission and the collection can be in two different locations. One could therefore operate the emitter in a low humidity environment which would have high atmospheric transmission for the emission and thus high cooling power, but place the dew collector in a warm, high humidity environment, like the basement of a building. In some respects, this was the avenue pursued by Zhou \textit{et al} who blew high humidity, high temperature air past their emitter \cite{zhou2018accelerating}.  

\section{Conclusions}
To conclude, in this paper we have discussed new theoretical limits for dew collection using black-body radiative cooling. We built an experiment to simulate the theoretical predictions that replicated a radiative emitter employing a gravity-fed, covered condensation surface. The experimental results were in excellent agreement with the theoretical predictions. This simple, passive design has no moving parts and greatly improves both maximum yields and the conditions under which dew collection can occur.  We believe there is a straightforward scaling of the system allowing for significant passive water production.  

\section{notes}
While working on this paper, we discovered similar work by Zhen \textit{et al} \cite{Zhen2019}.

\section{acknowledgements}
We would like to thank Joseph Shaw, Ronni Kozlov, Ziv Cohen, and Ahia Amrosi. We are grateful to the Hebrew University of Jerusalem for supporting this project.

\bibliography{DewCollection}

\end{document}